\documentclass[fleqn,usenatbib]{mnras}
\usepackage{newtxtext,newtxmath}
\usepackage[T1]{fontenc}
\usepackage{ae,aecompl}

\usepackage{graphicx}	% Including figure files
\usepackage{amsmath}	% Advanced maths commands
\usepackage{amssymb}	% Extra maths symbols
\usepackage[normalem]{ulem}

\title[Galactic bar torque]{Quantifying torque from the Milky Way bar using Gaia DR2}

\author[Kipper et al]{
Rain Kipper$^{1}$\thanks{E-mail: rain.kipper@ut.ee},
Peeter Tenjes$^{1}$,
Taavi Tuvikene$^{1}$,
Punyakoti Ganeshaiah Veena$^{1,2}$, \and
Elmo Tempel$^{1}$
\\
$^1$Tartu Observatory, University of Tartu, Observatooriumi 1, 61602 T\~oravere, Estonia\\
$^2$Kapteyn Astronomical Institute, University of Groningen,PO Box 800, 9747 AD Groningen, The Netherlands
}

\date{Accepted 2020 March 31. Received 2020 March 31; in original form 2019 October 14}

\pubyear{2020}

\begin{document}
\label{firstpage}
\pagerange{\pageref{firstpage}--\pageref{lastpage}}
\maketitle
\begin{abstract}
We determine the mass of the Milky Way bar and the torque it causes, using \textit{Gaia} DR2, by applying the orbital arc method. 
Based on this, we have found that the gravitational acceleration is not directed towards the centre of our Galaxy but a few degrees away from it. We propose that the tangential acceleration component is caused by the bar of the Galaxy. Calculations based on our model suggest that the torque experienced by the region around the Sun is $\approx 2400\,{\rm km^2\,s^{-2}}$ per solar mass. The mass estimate for the bar is $\sim 1.6\pm0.3\times10^{10}\,\mathrm{M_\odot}$. 
Using greatly improved data from \textit{Gaia} DR2, we have computed the acceleration field to great accuracy by adapting the oPDF method (Han et al. 2016) locally and used the phase space coordinates of $\sim 4\times10^5$ stars within a distance of 0.5 kpc from the Sun. {In the orbital arc method, the first step is to guess an acceleration field and then reconstruct the stellar orbits using this acceleration for all the stars within a specified region. Next, the stars are redistributed along orbits to check if the overall phase space distribution has changed. We repeat this process until we find an acceleration field that results in a new phase space distribution that is the same as the one that we started with; we have then recovered the true underlying acceleration.}
\end{abstract}


\section{Introduction} \label{sec:introduction}

\textit{Gaia} satellite data releases allow us to construct {quite} detailed models for the Milky Way (MW) stellar density distribution and its kinematics. The latest Data Release 2 \citep{Lindegren:2018} gives us an excellent opportunity to explore the solar neighbourhood (SN) and somewhat more distant regions. In the present paper, we calculate the gravitational acceleration of the MW using the \textit{Gaia} DR2 data, in an ellipsoidal region {within a} distance of 0.5 kpc from the Sun in the Galactic plane. 

{ Modelling the MW is very different from modelling other disc galaxies since we make observations from within the MW}. Although our location within the MW can make modelling easier, {(e.g. individual stars are resolved)} it can also add complications to it, {e.g. dust attenuation and selection function can have a strong influence on modelling}.
For example, it was only at beginning of the 1980s that the first direct hints that the MW may be a barred spiral galaxy came to light \citep{Matsumoto:1982}. This was possible because of near-IR observations. Due to dust attenuation and our position inside the MW, it was difficult to draw such a conclusion on the morphology of MW before that.

On the other hand, we are at a great advantage because of the wealth of observational data available for the MW, unmatched and unavailable for other galaxies. For instance, axisymmetric models developed by \citet{Piffl:2014, Mcmillan:2017, Binney:2017} use H\ion{I} and CO velocities, maser data, Sgr A* proper motions, {the globular cluster system, }the velocity distribution in the SN, SDSS star counts in different colours, RAVE data, detailed MW satellite data and N-body simulation data. Additional constraints on the mass distribution were derived from cold stellar streams \citep{Bovy:2016} and \textit{Gaia} DR2 proper motions of globular clusters \citep{Watkins:2019}. 
However, the assumption of axisymmetry in mass distribution models is only a first approximation. 

The existence of the central bar of the MW was first confirmed by \citet{Weiland:1994}, by analysing asymmetries in the near-IR surface brightness distribution of the central bulge from the COBE/DIRBE data. This was further confirmed by correcting the data for extinction  \citep{Dwek:1995, Binney:1997}.

There are currently two {contrasting scenarios} -- a fast rotating bar and a slow rotating bar. In the first case \citep[e.g.][]{Binney:1997, Bissantz:2003, Monari:2017} the bar is rotating with pattern speed $\Omega_p = 50 - 70~\mathrm{km\,s^{-1}kpc^{-1}}$; in the second case \citep{Wegg:2013, Wegg:2015, Portail:2015, Dias:2019} the calculated pattern speed is $25 - 30~\mathrm{km\,s^{-1}kpc^{-1}}$. Intermediate pattern speed values were derived by \citet{Li:2016, Portail:2017, Perez:2017, Sanders:2019, Bovy:2019} as $\Omega_p = 35 - 40~\mathrm{km\,s^{-1}kpc^{-1}}$. 
These calculated pattern speeds vary by about two times and as a result their corotation radii and outer Lindblad radii vary quite significantly. Both these scenarios agree that the angle between the major axis of the bar and the line connecting the Sun and the Galactic centre (GC) is about $20 - 30\,\mathrm{deg}$.

According to the axisymmetric models, the stars in orbits are phase-mixed. According to the non-axisymmetric models, stellar orbits are somewhat perturbed and phases of stars in orbits may not be completely mixed \citep{Dehnen:2000, Fux:2001, Monari:2017, Binney:2018, Trick:2019} and thus orbital structure is more complicated {(i.e. there are resonances)}.
For example, using the \textit{Gaia} DR2 data, \citet{Ramos:2018} found that, in the case of the MW, some orbit phases are mixed. Similar arcs and ridges were also found by \citet{Antoja:2018, Kawata:2018, Trick:2019}. Gravitational potential disturbances due to the bar may have caused deviations of stars from their initial orbits in the case of several cold stellar streams \citep{Hattori:2016, Pearson:2017, Banik:2019} {that originate from small stellar systems}. The torque from the bar is not the only reason (see e.g. \citet{Kipper:streams}). 
{These disturbances can }create observed gaps in stream surface density distributions. 

Unfortunately, the structural parameters of the bar and its contribution to the gravitational acceleration are still rather poorly constrained. Thus, it is important to know the gravitational acceleration distribution in the Galactic plane and also to study how this allows one to constrain the bar properties. The \textit{Gaia} satellite data provide an excellent opportunity to do this. 

In the present paper we calculate all three acceleration components in the SN. We use the orbital arc method, developed in \citet{Kipper:2019}. The method and its specific implementation details are described in Sect.~\ref{sec:method}. The method is used on the \textit{Gaia} DR2 data. We use two different versions of the data, from the StarHorse project \citep{Khalatyan:2019} and from the Sch\"onrich catalogue \citep{Schonrich:2019}. The selection of the  data used is described in Sect.~\ref{sec:data_and_selection}. We demonstrate in Sect.~\ref{sec:results} that the derived acceleration components cannot be explained within an axisymmetric model. The final section is dedicated to the summary and discussion. 

We denote $(x, y, z)$ as Galactocentric rectangular coordinates and $(R, \theta , z)$ as corresponding cylindrical coordinates, where $\theta = 0$ corresponds to the opposite direction {from} the Sun. Transformations of sky coordinates, proper motions and radial velocities to Galactocentric coordinates and velocities were carried out using the Astropy package \citep{astropy:2013,astropy:2018}. For the solar velocity, we used the values $(U_\odot, V_\odot, W_\odot) = (11.1,12.24,7.25)\,\mathrm{km}\,\mathrm{s}^{-1}$, and $V_{g,\odot} = V_{c,\odot} + V_\odot$, with the circular velocity $V_{c,\odot} = 240\,\mathrm{km}\,\mathrm{s}^{-1}$ \citep{Lopez-Corredoira:2019}. 
\begin{figure}
    \centering
    \includegraphics[width=\columnwidth]{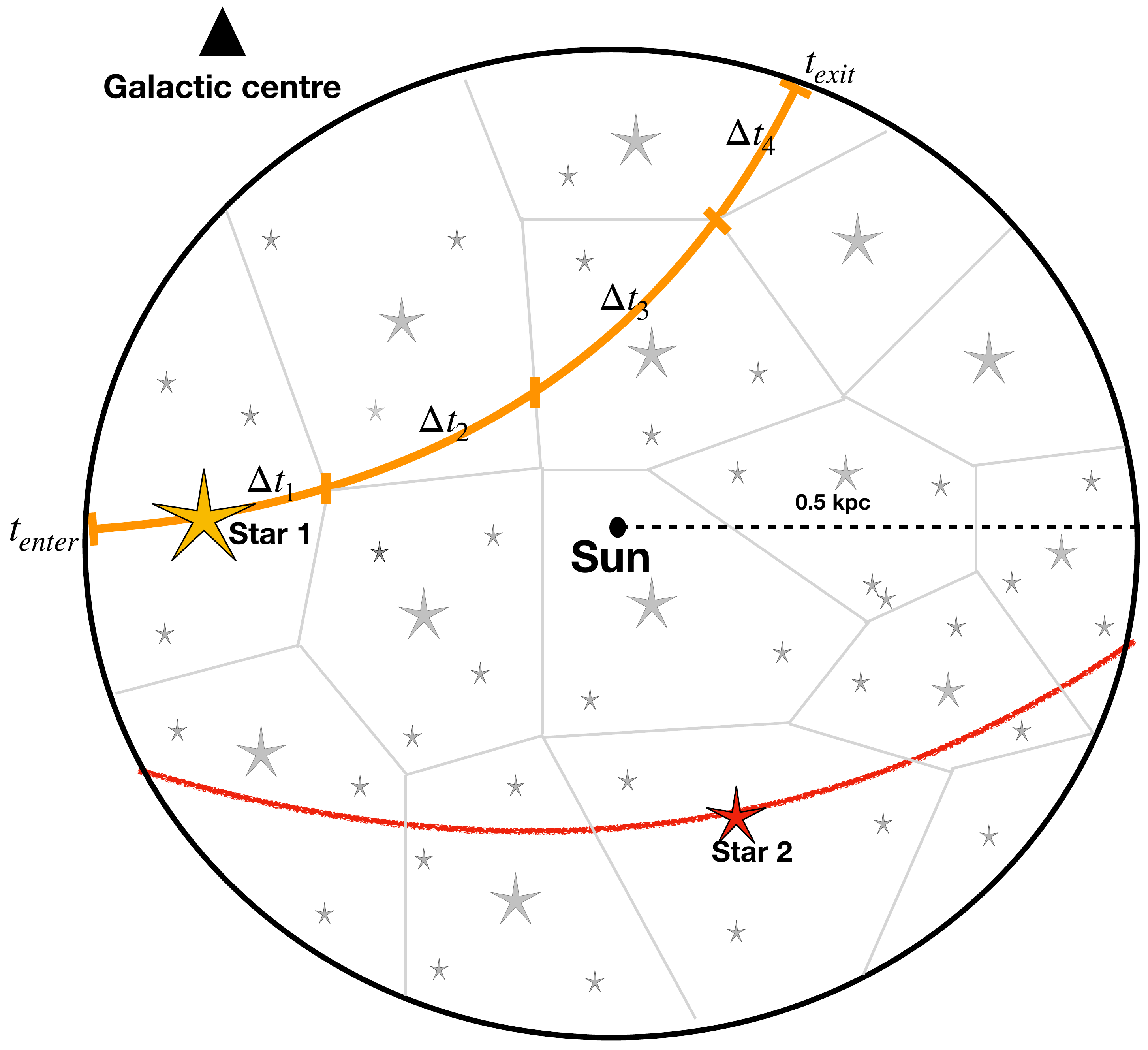}
    \caption{An illustration of the region where the orbital arc method is applied. The central point represents the Sun and the circular region up to a distance of 0.5 kpc from the Sun is the region used in this paper. The black triangle {points towards the Galactic centre}. The coloured arcs for star 1 and star 2 represent the reconstructed orbits for these two stars. Orbital arcs are reconstructed for all stars in this region. The grey cells are the Voronoi cells, each of which contains a similar number of stars. The time interval, $\Delta t_i $, is the time a star spends in the $i^\mathrm{th}$ Voronoi cell. The times at which the star enters and exits the region are $t_{\rm enter}$ and $t_{\rm exit}$, respectively. }\label{fig:voronoiDiagram}
    \vspace{-0.2cm}
\end{figure}

\section{Method and implementation}\label{sec:method}
\subsection{Orbital arc method}\label{sec:four_steps}
{In this section, we provide a brief overview of the orbital arc method, which we have implemented in this paper to compute the gravitational acceleration, mass and torque estimates of the Galactic bar. For a detailed and thorough description of the formulation and tests of the model, please see \citet{Kipper:2019}. We will refer to this particular method as the orbital arc method, since its most crucial step is the reconstruction of stellar orbits to accurately obtain the acceleration in the Milky Way, using the phase space information of stars. }
{This has already been successfully applied to a simulation in \citet{Kipper:2019} and for the observational data in a simplified form \citep{Kipper:2018}. Here we apply it for the \textit{Gaia} DR2 data.}
{The orbital arc method
has five important steps. 
\begin{enumerate}
    \item[{\textbf{Step 1 -- Acceleration field:}}] We first select a region with a sufficiently large number of stars and known phase space coordinates. Next, we guess an acceleration field and use this to get the orbits. In the orbital arc method, the acceleration field is a free function of the model and contains free parameters. 
    For instance, we can take advantage of the axisymmetric property of a galaxy, and choose an acceleration field described by the cylindrical coordinates:
\begin{eqnarray}
    a_x &=& a_R\cos\theta ,  \label{eq:form_mustbebar1}\\
    a_y &=& a_R\sin\theta + A_y ,  \label{eq:form_mustbebar2}\\
    a_z &=& a_z. \label{eq:form_mustbebar2b}
\end{eqnarray}  
Components $a_R$ and $a_z$ are taken in the form of functions 
\begin{eqnarray}    
    a_z &=& A_z + A_{z,z}z + A_{z,R}\Delta R + A_{z,Rz}z\Delta R , \label{eq:form_mustbebar3}\\
    a_R &=& A_R + A_{R,R}\Delta R ,  \label{eq:form_mustbebar4}\\
    \Delta R &=& R-R_\odot,\label{eq:form_mustbebar5}
\end{eqnarray}
where $A_y$, $A_z$, $A_R$, $A_{z,z}$, $A_{R,R}$, $A_{z,R}$, $A_{z,Rz}$ and in some cases also $R_\odot$ are free parameters obtained via fitting.

If we do not assume axisymmetry, acceleration vector components are taken in the form of their {first-order} Taylor expansion:
\begin{eqnarray}
    a_x = A_x + A_{x,x}\Delta x + A_{x,y}\Delta y + A_{x,z}\Delta z\label{eq:form_cart_1} \\
    a_y = A_x + A_{y,x}\Delta x + A_{y,y}\Delta y + A_{y,z}\Delta z\label{eq:form_cart_2} \\
    a_z = A_x + A_{z,x}\Delta x + A_{z,y}\Delta y + A_{z,z}\Delta z\label{eq:form_cart_3}. 
\end{eqnarray}
Here $\Delta x$, $\Delta y$, $\Delta z$ denote the distances from the region's centre.
    \item {\textbf{Step 2 -- Orbital arc reconstruction:}} Using the initial conditions, which is the phase space information from the data, and the acceleration field from the previous step, we solve the equations of motion to obtain stellar orbits for each star. A schematic of the reconstructed orbit arcs is represented as coloured arcs in Fig.~\ref{fig:voronoiDiagram}. 
    \item {\textbf{Step 3 -- Randomizing star position:}} The core of the proposed method lies in the oPDF (orbital Probability Density Function), according to which the time of observing a star is random. This means, we can reposition a star along its orbit by picking a random time from a uniform distribution of time. By picking infinitely many times from this distribution, we reach a continuous distribution of the star along its orbit (a similar description to the procedure can be found in \citealt{Han:2016}). Following this, we reposition every star in its orbit and get a new distribution of stars in the region. This is relevant in the last step where we will compare the old and  new distributions. 
    \item {\textbf{Step 4 -- Phase space density:} } In order to compute the probability of finding a star in its orbit, we need to first specify a small segment of the star's orbit. For this, we construct  Voronoi tessellations by considering small subsets of data, such that in each Voronoi cell there are similar numbers of stars; this reduces the Poisson error. The Voronoi cells are shown in the schematic diagram in Fig.~\ref{fig:voronoiDiagram}. To compute the probability of finding a star in its orbit, we use the Voronoi cells as the orbital segments. For example, for star~1 in Fig.~\ref{fig:voronoiDiagram}, the time spent by the star in each Voronoi cell along its orbit is given as $\Delta t_i$. So, the probability of finding that star in the $i^\mathrm{th}$ Voronoi cell is the time spent in that Voronoi cell, $\Delta t_i$, divided by the time spent in the entire region, which is, ${\Delta t_i}/{\left(t_{\rm exit} - t_{\rm enter}\right)}$ as seen in Fig.~\ref{fig:voronoiDiagram}. Eventually, a combined probability is calculated for each Voronoi cell, which is the sum of probabilities of all stars in each cell.  
    \item {\textbf{Step 5 -- Comparing phase space density distributions:}} In this final step we compare the phase space distribution of the original data and the phase space distribution of the newly positioned stars. The phase space distribution comparison is done statistically by computing the likelihood. If the likelihood is not maximum then the entire process is repeated with new acceleration field parameters. 
    
    Eventually, the orbital arc method will give the acceleration field corresponding to the maximum likelihood, which is the field that describes the true underlying acceleration of the MW. The distribution of likelihoods gives the statistical uncertainty. 
\end{enumerate}
}
{Since the level of accuracy relies on the available data, we need the phase space coordinates of a sufficiently large number of stars. Hence, \textit{Gaia} DR2 is aptly suited for the study.}

\subsection{Implementation: the smoothing kernel}\label{sec:impl_grid}
{In order to compare the phase space distributions of the original data and the repositioned stars, we have used Voronoi cells to get a smooth phase space density. This is achieved by computing the time stars spend in each Voronoi cell, as described in step 4 in Sect.~\ref{sec:four_steps}. In Fig.~\ref{fig:voronoiDiagram}, $\Delta t$ represents the time a star spends in a cell. The ratio of the time spent by a star in a Voronoi cell, $\Delta t_i$, to the time that it spends in the entire region (see $t_{\rm enter}$ and $t_{\rm exit}$ in Fig.~\ref{fig:voronoiDiagram}) gives the phase space density of this model.}

The shapes and sizes of the regions in which we intend to calculate the  accelerations are mainly motivated by the quality of 6D data. The shapes of these regions and the Voronoi cells used to smooth phase space\footnote{The Voronoi tessellation of the region is done in order to compare the original and the new phase space distributions.} should be complementary to each other. {For example, if the available data are of a spherical region, then a rectangular grid or cell is not the most optimal.} Therefore, the best possible grid should coarsely follow the distribution of data. One of the best ways to achieve this is by Voronoi tessellations, and we have hence used this method for the paper. However, in principle, any similar grid can be used. We used a random small subset of the data of about $\sim 100$ stars to obtain the {Voronoi cells}. 

Each grid-cell is described by two numbers: the closest tessellation centre in ordinary space and the closest tessellation centre in velocity space. These two indices are required to avoid combining velocity and distance data into a single quantity, because this kind of combination produces an additional free parameter that we wish to avoid. For example, by using $100$ data points to tessellate into a grid, we will have $100^2 = 10\,000$ independent cells, which is usually sufficient to describe the phase space distribution of about $\approx 420\,000$ stars (i.e. $42$ stars per cell). For the current study, we have selected $100$ cells  for each of position and velocity space, unless noted otherwise.

\subsection{Implementation: flux limitedness}\label{sec:impl_fluxlim}
{Flux-limited observational data are a natural constraint in large surveys.}
There are two common approaches to overcome {this}: a) to construct a volume-limited sample and discard some data, or b) to use all the data and add a weight to each point. 

{Most dynamical modelling methods are constructed based on the assumption that we are able to observe everything, i.e. the volume-limited approach. Some specifics of the present modelling allow us to use the advantage of increased amounts of data of the flux-limited selection, while essentially using the method constructed for the volume-limited approach. This approach is described further in this section. }

A volume-limited selection is one in which both the stellar distance from an observer and absolute magnitudes of its stars are constrained by a flux limit of the sample. This grants that all of the stars would remain observable if we randomize their position in the region. {Our aim is to combine volume-limited selections to acquire methodology that allows us to use flux-limited data.} Let us denote $m_{\rm lim}$ as the completeness limit of the flux-limited sample. Then the corresponding absolute-magnitude limit $M_{\rm lim}$ and the distance limit $d_{\rm lim}$ are related by $5\log_{10}d_{\rm lim} = m_{\rm lim} - M_{\rm lim} + 5$ (at the moment we ignore the attenuation correction). It is possible to construct a volume-limited sample by selecting only stars that have $M < M_{\rm lim}$ and $d < d_{\rm lim}$. The same applies if we use an additional cut from higher absolute magnitudes, leaving $M$ in the range $M_{\rm lim}-\Delta M < M \leq M_{\rm lim}$. Following the denotations from \citet{Kipper:2019}, the observed phase space density as a function of phase space coordinates (${\bf q}$) for a volume-limited sample can now be written as $p_{\rm obs}({\bf q}|M_{\rm lim},d_{\rm lim})$, and the model one as $p({\bf q}|M_{\rm lim},d_{\rm lim},\zeta)$. Here $M_{\rm lim}$ and $d_{\rm lim}$ are not independent, but tied to the absolute-magnitude definition. For the correct gravitational acceleration parameters $\zeta$, and irrespective of the absolute-magnitude limit, these distributions must match: 
\begin{equation}
    p({\bf q}|M_{\rm lim},d_{\rm lim},{\bf \zeta}) = p_{\rm obs}({\bf q}|M_{\rm lim},d_{\rm lim}).
\end{equation}
If this relation matches for each small volume-limited sample, then the relation must hold for the sum (or integral) of these small volume-limited samples as well:
\begin{eqnarray}
    \int p({\bf q}|M_{\rm lim},d_{\rm lim}(M_{\rm lim}),{\bf \zeta})\, {\rm d}M_{\rm lim} =\nonumber \\ 
    = \int p_{\rm obs}({\bf q}|M_{\rm lim},d_{\rm lim}(M_{\rm lim}))\, {\rm d}M_{\rm lim}.
\end{eqnarray}
This means that we may integrate an orbit until the apparent magnitude of the corresponding star reaches the limiting magnitude $m_{\rm lim}$ due to its changing distance, and smooth the position of the star along its orbit within that limit. In Sect.~\ref{sec:test_fluxlim} we test the validity of this approach. 

\subsection{Requirements for data}
{An integral part of the method is orbit calculation. This has two ingredients: the proposed acceleration function and initial conditions for the orbits. As an analytical expression the first one is infinitely precise for each likelihood evaluation. The second one is as precise as the data allow. Imprecisions in the data are amplified by the orbit integration, i.e. $\Delta x \sim \Delta x_0 + t\Delta v$, where $\Delta$ denotes uncertainty for positions and velocities respectively. This shows that the uncertainties accumulate with time; hence the position of a star is unknown in some cone. {Due to  uncertainties (especially heteroscedastic ones) in the \textit{Gaia} data combined with smoothing phase space, we may reconstruct imprecise orbits, which will introduce a bias in the acceleration.} The simplest way to avoid these problems is to use maximally precise data. }

{The second requirement is to have a sufficient amount of data. This is needed to describe the phase space density sufficiently precisely. Assuming that the data are very precise, the only source of uncertainty is the Poisson noise from the sampling. }

\section{Observational data}\label{sec:data_and_selection}

\subsection{Construction of the data sample}\label{sec:data}

Six-dimensional high-quality phase space coordinates in the SN are now available from \textit{Gaia} satellite Data Release 2 for a significant number of stars. At present there are three catalogues available based on the \textit{Gaia} measurements {and including estimated star distances}: the Gaia Collaboration catalogue \citep{Lindegren:2018}, the StarHorse project catalog, SH, \citep{Khalatyan:2019} and the Sch\"onrich catalog, Sc, \citep{Schonrich:2019}. There is a known issue concerning the zero point of parallaxes from the Gaia Collaboration, which is overcome in the latter two catalogues. Therefore we selected these two catalogues as our main sources of input data and calculated our results for both of them separately. 
To calculate gravitational acceleration, we need to know mass density gradients. Although the main source of density gradients results from the smooth density distribution of the MW, selection effects can produce artificial gradients. The two dominant ingredients for this kind of selection effects are Malmquist bias (covered in the previous section) and dust attenuation. To suppress the effects from dust attenuation, we use 2MASS \citep{Skrutskie:2006} catalogue $J$-band magnitudes where extinction is negligible. 

\begin{figure}
    \centering
    \includegraphics[width=\columnwidth]{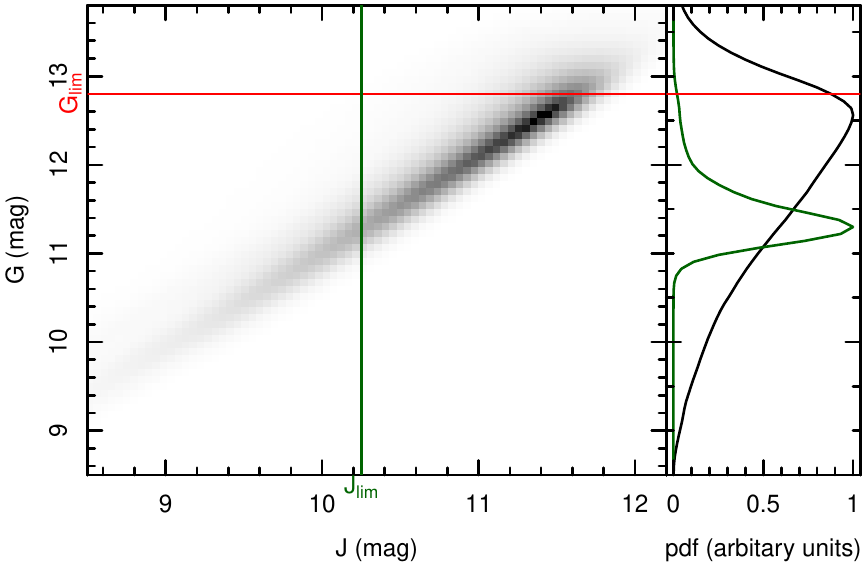}
    \caption{Distribution of apparent magnitudes of all stars within $0.5$~kpc from the Sun. The $G$-band magnitudes are from the \textit{Gaia} data and the $J$-band magnitudes are from the  2MASS survey. The red horizontal line and the green vertical line depict the spectroscopic completeness limit and the limiting magnitude $J_{\rm lim}$ respectively of our main sample. The right-hand panel shows the distribution of all stars from \textit{Gaia}. The black line shows all the stars and the green line shows only those with {magnitudes} brighter than $J_{\rm lim}$. The distribution of our sample of stars drops before reaching the \textit{Gaia} completeness limit. Only a fraction of $0.0008$ stars have a higher $G$-band magnitude; therefore we choose our sample based on 2MASS {photometry}. }\label{fig:valim_gj}
\end{figure}

The cross-match between the \citet{Khalatyan:2019} SH and 2MASS catalogues gave 6\,964\,515 entries; between the \citet{Schonrich:2019} Sc and 2MASS catalogues there were 6\,519\,209 matches. We constrained the input {magnitudes} in such a way that the \textit{Gaia} $G$-band completeness (being affected by dust attenuation) has substantially less effect than our selection based on the $J$ {passband} (being nearly attenuation free), i.e. $P(G>G_{\rm lim}|J<J_{\rm lim}) \ll 1$. The {apparent-magnitude} data within $0.5~{\rm kpc}$ from the Sun for our selected sample are shown in Fig.~\ref{fig:valim_gj}. A strong correlation between the $G$- and $J$-band magnitudes catches the eye. The $J$-band limit $J_{\rm lim}$ was fixed to a value where the distribution of brighter stars in the $G$ band ends before reaching the \textit{Gaia} spectroscopic completeness limit $G_{\rm lim}$. This is shown as a green line in the left-hand panel and the corresponding probability density distribution $p(G|J<J_{\rm lim}){\rm d}G$ on the right-hand panel. The fraction of $G$ magnitudes crossing  $G_{\rm lim}$ is $8\times 10^{-4}$ for the adopted $J_{\rm lim}=10.25$; hence we conclude that our sample is almost independent of the \textit{Gaia} completeness limit and dust attenuation. 

The smooth acceleration distribution of the MW is taken as an input in modelling process and it does not include local potential wells of stellar clusters. Hence, we cannot describe the motion of stars within clusters and must exclude these cluster stars from our sample. We excluded all stars that appeared to be cluster members in catalogues by \citet{Cantat-Gaudin:2018} or the \citet{Gaia:2018:HR}. In total, $993$ stars or about $0.2$ per cent of stars from the final selection were excluded.

\subsection{Selection of the region}\label{sec:region}
In the paper where we presented the method and tested it on simulation data \citep{Kipper:2019}, we aimed to use rather small regions in order to have a simple analytical form for acceleration vector components. In the current paper, we selected a larger region to suppress Poisson noise and to increase the region size in the radial direction to have a stronger basis to also estimate the first derivative of the radial acceleration. This changes our approach somewhat: instead of using a simple form for accelerations, we now try to model the underlying acceleration field with a well-motivated analytical form. 

Thus, due to available data, instead of using several small regions {as we did in \citet{Kipper:2019}}, we selected one larger region, {as shown in the schematic in Fig.~\ref{fig:voronoiDiagram}}. Our main aim was to recover the acceleration field in the plane of the Galaxy; hence we constructed a region where accelerations in the MW plane have a longer time to act on stars. In the vertical direction, density gradients are much steeper and one may expect that the corresponding accelerations may have also more complex forms. To avoid using more complex accelerations in vertical directions, we selected a thin region. 

{The region size is selected to balance two previous effects: maximally small to have a simple acceleration form and maximally close to keep the observational uncertainties low, and at the same time maximally large to let acceleration act for a sufficiently long time in the model. }
The boundary of the selected region is described by a biaxial ellipsoid:
\begin{equation}
    \left(\frac{\Delta x}{x_{\rm max}}\right)^2 + \left(\frac{\Delta y}{y_{\rm max}}\right)^2 + \left(\frac{\Delta z}{z_{\rm max}}\right)^2 = 1, \label{eq:ell}
\end{equation}
with $x_{\rm max} = y_{\rm max} = 0.494\,{\rm kpc}$ and $z_{\rm max} = 0.218\,{\rm kpc}$. Here $\Delta x$, $\Delta y$ and $\Delta z$ denote the coordinates from the centre of the region. The centre of the region is at $(x, y, z) = (-8.3, 0.0, 0.0)$ in kpc. The position of the Sun with respect to the region centre is $(-0.040, 0.0, 0.027)$ in kpc. Within this region there are $417\,727$ stars when using the \citet{Schonrich:2019} catalogue (Sc), and $426\,767$ stars when using the StarHorse (SH) catalogue. {Larger regions would require the use of a precise selection function, which would complicate the analysis.} 

\section{Results}\label{sec:results}
{We calculated gravitational acceleration in the region around the SN as described in Sect.~\ref{sec:region} using the method and its implementation explained in Sect.~\ref{sec:method}. In order to decipher various aspects of the acceleration field (e.g. deviations from axisymmetry), we used different functional forms to describe the underlying acceleration. }

\subsection{Calculated acceleration components}\label{sec:cartesian}
In our first attempt to model the acceleration in the region we did not specify a design-based form of an overall gravitational potential of the MW. Instead we assumed that any acceleration form can be approximated with their Taylor expansions Eqs.~\eqref{eq:form_cart_1} -- \eqref{eq:form_cart_3} and we fit the coefficients of this acceleration ($A_x$, $A_{x,x}$, $A_{x,y}$, $A_{x,z}$, $A_y$, $A_{y,x}$, $A_{y,y}$, $A_{y,z}$, $A_z$, $A_{z,x}$, $A_{z,y}$, $A_{z,z}$). This way of modelling is powerful because it allows us to not only model the acceleration in a tiny region, but in principle the entire MW if we can get the overall gravitational potential. We fit a total of $12$ free parameters, $A_i$ and $A_{i,j}$ for the flux-limited samples of stars within the selected region (see Sect.~\ref{sec:region}) for both the Sc and SH catalogues of \textit{Gaia} DR2 (for more details see Sect.~\ref{sec:data}).

We used $100$ random points to describe the grid; hence, there are $\approx 42$ stars per grid bin. The fitting was done with the \textit{Multinest} code \citep{MN1, MN2, MN3} using $500$ live points. To include the randomness caused by the gridding, we ran the code eight times and averaged the posterior distribution of different runs. All the modelling was done in this way, unless noted otherwise. The priors of the Bayesian modelling were chosen to be of uniform distribution with the limiting values provided in Table~\ref{tab:cartesian}. In the table we give the posterior distribution of each parameter $\zeta$ with five quantiles positioned at $P(\zeta) = \{0.023, 0.159, 0.500, 0.841, 0.977\}$.  

Previous studies have shown that the Sun is not located precisely at the centre of the Galactic plane, but is about $25\,{\rm pc}$ away from it \citep{BlandHawthorn:2016}. Thus, there must be an acceleration component in the vertical direction, as confirmed in e.g. \citet{Kipper:2018} based on dynamics. In Table~\ref{tab:cartesian} our estimation of the vertical acceleration $A_z$ and its gradient in the $z$-direction $A_{z,z}$ are given. Using these two values and by making a linear approximation at distances close to the plane, we deduce that we are located at $ z_\odot\approx A_z/A_{z,z} = \{ -111^{+33}_{-76} ({\rm SH}),\, -117^{+58}_{-66} ({\rm Sc})\} $~pc from the vertical coordinate value defined as having zero vertical acceleration in contrast to the symmetry-defined centre. 

\begin{figure}
    \centering
    \includegraphics[width=\columnwidth]{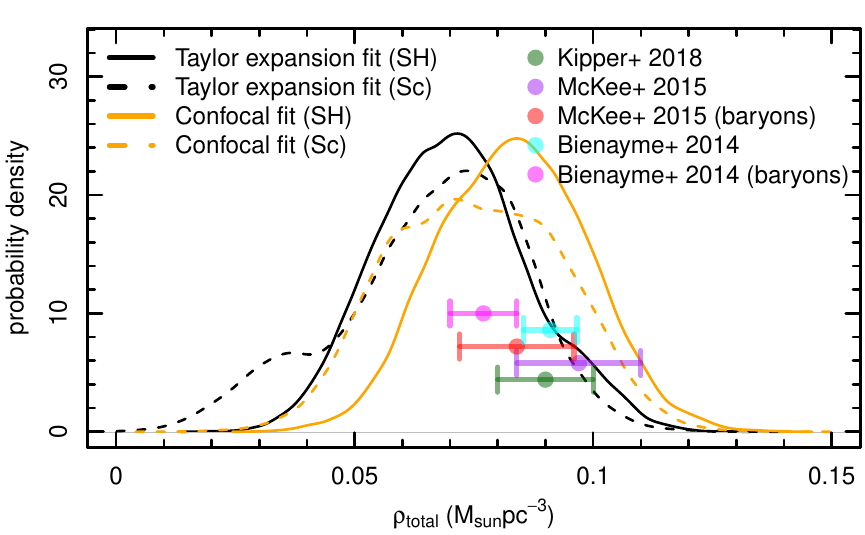}
    \caption{The figure shows the average matter density in the solar neighbourhood and is compared with the results from \citet{Bienayme:2014, McKee:2015, Kipper:2018}. These results do not match very well because they use different datasets and different assumptions of the underlying acceleration. The high uncertainty in the calculated results is due to the optimization of the selected acceleration form to determine accelerations in Galactic plane. Note that this is not the vertical component, which is usually used to determine the overall matter density.}
    \label{fig:total_den_cart}
\end{figure}

The Poisson equation combines the gravitational potential and total mass density. By using calculated acceleration components in the Poisson equation, we can compute the average total matter density in our selected region as:
\begin{equation}
    \nabla^2\Phi = -A_{x,x} - A_{y,y} - A_{z,z} = 4\pi G\rho_{\rm total}, \label{eq:poisson_cartesian}
\end{equation}
where $\Phi$ is the gravitational potential and $\rho_{\rm total}$ is the total mass density inside the region. Calculated Taylor expansion fit components for two different \textit{Gaia} catalogues give us 
$\rho_{\rm total} = 0.070^{+0.016}_{-0.016}{\rm \,M_\odot\,pc^{-3}}$ (SH catalogue) and $\rho_{\rm total} = 0.069^{+0.016}_{-0.023} {\rm \,M_\odot\,pc^{-3}} $
(Sc catalogue).
The full probability density distribution of the total matter density $\rho_{\rm total}$ can be seen in Fig.~\ref{fig:total_den_cart}. One must bear in mind, that this total density value applies as the average in this region (extent in the $z$-direction is $2\times0.22$ kpc). In order to describe the changes in the vertical component of the acceleration, we need a more sophisticated form of acceleration as the linear form cannot capture quick density changes along the $z$-direction.
Therefore, if we use another form by assuming axisymmetry with respect to the centre, then the number of free parameters can be significantly reduced and more concrete conclusions can be made. Selected Taylor expansion might not fully grasp all the details of the vertical structure, since it is described with just one free parameter ($A_{z,z}$).

\subsection{Deviations from a simple axisymmetric MW model} \label{sec:mustbebar}

\begin{figure}
    \centering
    \includegraphics[width=\columnwidth]{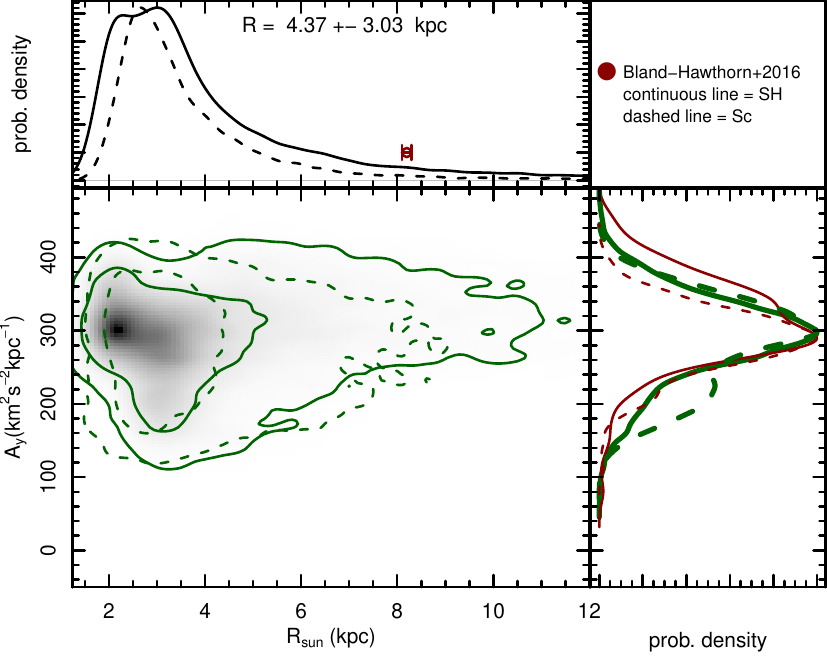}
    \caption{The \textit{central panel} shows the correlation between the centre of acceleration, $R_\odot$, and the component of the acceleration vector, $A_y$, as described in Sec.~\ref{sec:mustbebar}. The \textit{top panel} shows the distribution of $R_\odot$ for an axisymmetric fit. The brown point at  $8.3~$kpc shows the distance of the Sun to the centre of our Galaxy. This indicates that the curvature of the isopotential lines is very likely less than $8.3~$kpc. The \textit{right-hand panel} shows the distribution of the $A_y$ component of the acceleration vector at the region centre. The green lines are for the overall posterior distribution and the red lines are for the subset where $R_\odot\approx8.3\,{\rm kpc}$. This figure highlights the necessity to include the non-axisymmetric component to fit the acceleration, since the default for the MW at $(R_\odot = 8.3, A_y=0)$ does not account for what is observed. 
     }
    \label{fig:mustbebar}
\end{figure}

In case of a stationary axisymmetric MW, the acceleration component along the direction of Galactic rotation $a_y(\Delta x, \Delta y=0, \Delta z) = 0$ and equipotential curves are concentric circles. The median values of acceleration computed within the selected region using the coordinates $(x, y, z) = (-8.3, 0, 0)$~kpc as the centre are 306 and $284\,{\rm km^2\,s^{-2}\,kpc^{-1}}$ for the SH and Sc catalogues respectively. Results of these calculations along with $1\sigma$ and $2\sigma$ limits are shown in Fig.~\ref{fig:mustbebar} and the used priors are given in Table \ref{tab:mustbebar}. They are designated as 'Sc, flux' and 'SH, flux'. None of the $27\,748$ posterior samples from \textit{multinest} show negative $A_y$ values. Thus, the results are not consistent with axisymmetry.

{Based on the assumption that equipotential curves are concentric circles, we derived the radius of this circle. The radii are $3.4\,{\rm kpc}$ for the SH catalogue and $ 3.2\,{\rm kpc}$ for the Sc catalogue.}
Most of the posterior distribution had values lower than $8.3$~kpc, i.e.  $P(R_\odot~>~8.3\,{\rm kpc})~=~\{ 0.048 ({\rm Sc}),\, 0.11 ({\rm SH}) \}$. Hence, the 'acceleration centre' is most likely closer to us than the GC and equipotential curves have higher curvature than one would expect for the distance to the Galactic centre. Thus we conclude that axisymmetric potential distribution is not valid at SN, and interpret it as an argument to support the presence of a rather massive central bar. 

As already explained in Sect.~\ref{sec:test_fluxlim}, we calculated the { acceleration components} assuming axisymmetry, by selecting the components $a_R, a_z$ to be in the form of Eqs.~\eqref{eq:form_mustbebar3} -- \eqref{eq:form_mustbebar5}.
During the fitting the solar distance $R_\odot$ was also taken as a free parameter. Taking the posterior in these fits close to $R_\odot=8.3\,\mathrm{kpc}$, we found the radial acceleration to be $-6190^{+70}_{-160}\,{\rm km^2\,s^{-2}\,kpc^{-1}}$, which corresponds to the circular velocity 227~km\,s$^{-1}$. The combined estimate of the observed circular velocity at 8.3~kpc is somewhat larger, being $238\pm15~{\rm km\,s^{-1}}$ \citep{BlandHawthorn:2016}, but it is consistent with the calculated result within errors. 

\subsection{ Deriving the properties of the bar}
\label{sec:confocal}
To calculate the total mass of the bar, we assumed that spatial density distribution of the bar has the same form as that derived by \citet{Wegg:2015}:
\begin{eqnarray}
        \rho &=& \frac{M_{\rm bar}}{4\pi x_0y_0z_0}\exp{\left( -\left[ \left(\frac{x}{x_0}\right)^{c_\perp}+\left(\frac{y}{y_0}\right)^{c_\perp} \right]^{1/c_\perp} \right)}\nonumber\times\\
        &\times&\exp{\left[-\frac{z}{z_0}\right]} 
         {\rm Cut}\left[\frac{R-R_{\rm out}}{\sigma_{\rm out}}\right] {\rm Cut}\left[\frac{R_{\rm in}-R}{\sigma_{\rm in}}\right]\\
        {\rm Cut}(x) &=& \begin{cases}
             \exp(-x^2)& {\rm if\,}x>1 \\
             1& {\rm if\,}x\le 1.
        \end{cases}
\end{eqnarray} 
The values of the parameters $x_0,y_0,z_0,\sigma_{\rm in},\sigma_{\rm out},c_\perp,R_{\rm in}, R_{\rm out}$ were also taken to be the same as those derived by \citet{Wegg:2015}. 
In this form \citet{Wegg:2015} excluded symmetric parts of the Galaxy (e.g. bulge) by cut-off. By calculating tangential accelerations for this bar and fitting the calculated values with the values derived by us and referred to in the previous subsection, we derived that the mass of the bar (using the cut-off in previous equations) will be $0.41^{+0.07}_{-0.08} 10^{10}\mathrm{M}_\odot$ for the SH catalogue and $0.40^{+0.07}_{-0.11} 10^{10}\mathrm{M}_\odot$ for the Sc catalogue. 
Without the cut-off, by adopting their profile (and inferring their $M_{\rm bar}$) the overall bar mass would be 
$\{ 1.59^{+0.27}_{-0.31}({\rm SH}),\, 1.55^{+0.29}_{-0.43}({\rm Sc}) \}\,10^{10}\mathrm{M}_\odot$. 

In the previous subsection we concluded that the assumption that equipotential curves are circles is clearly not valid. Therefore, we assumed that these curves are confocal ellipses in the Galactic plane and then computed the acceleration due to the bar. An analytical form for accelerations describing the potential of the bar as confocal ellipsoids was chosen to be 
\begin{eqnarray}
    a_x &=& a_R\cos\theta + A_{R,{\rm bar}}\Delta x(\Delta x',\Delta y')\label{eq:focus_eq1}\\
    a_y &=& a_R\sin\theta + A_{R,{\rm bar}}\Delta y(\Delta x',\Delta y')\label{eq:focus_eq2}\\
    a_z &=& A_z + A_{z,z}z + A_{z,R}\Delta R + A_{z,Rz}z\Delta R \label{eq:focus_eq3},
\end{eqnarray}
where $a_R$ and $\Delta R$ are given by Eqs.~\eqref{eq:form_mustbebar4} and \eqref{eq:form_mustbebar5}. The normalized vector components of the potential gradient of the confocal bar, $\Delta x'$ and $\Delta y'$, are described by
\begin{equation}
    \frac{\Delta x'^2}{a^2} + \frac{\Delta y'^2}{a^2 - L_{\rm bar}^2} = 1.
\end{equation}
$L_{\rm bar}$ is the focal length of the equipotential curves and $a$ describes the size of the ellipsoid. {The coordinate transformation from $(x,y)$ to $(x',y')$ is done by rotating the original axes by an angle of $29.5^\circ$ which is the position angle of the major axis of the bar \citep{Wegg:2015}.}
Results from these calculations are given in Table~\ref{tab:my_label}, they contain the coefficients obtained from the fits.
\begin{figure}
    \centering
    \includegraphics[width=\columnwidth]{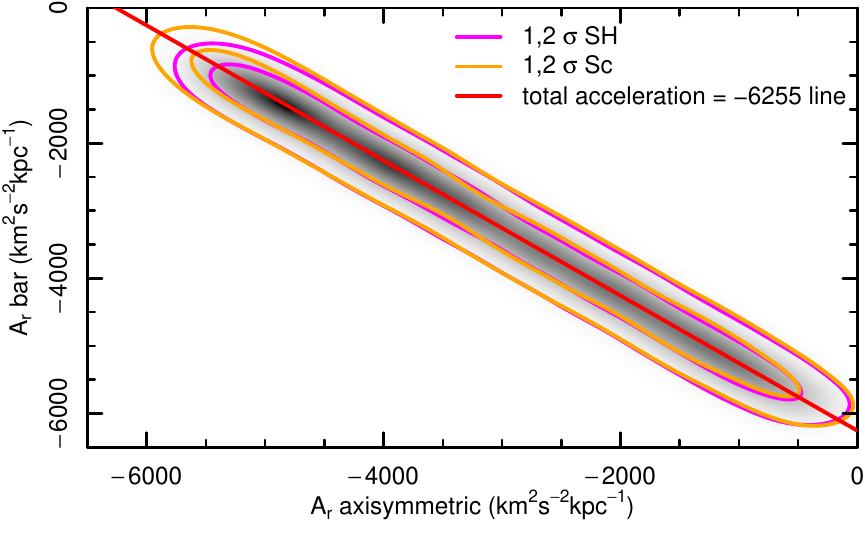}
    \caption{The relation between acceleration from the axisymmetric component and from the bar component. The contours show $1\sigma$ and $2\sigma$ confidence intervals for the \citet{Khalatyan:2019} (SH) and \citet{Schonrich:2019} (Sc) datasets. The strong correlation between them shows degeneracy of accelerations in the functional form of equations \eqref{eq:focus_eq1}-\eqref{eq:focus_eq3}.}
    \label{fig:ar_vs_ar_bar}
\end{figure}
\begin{figure}
    \centering
    \includegraphics[width=\columnwidth]{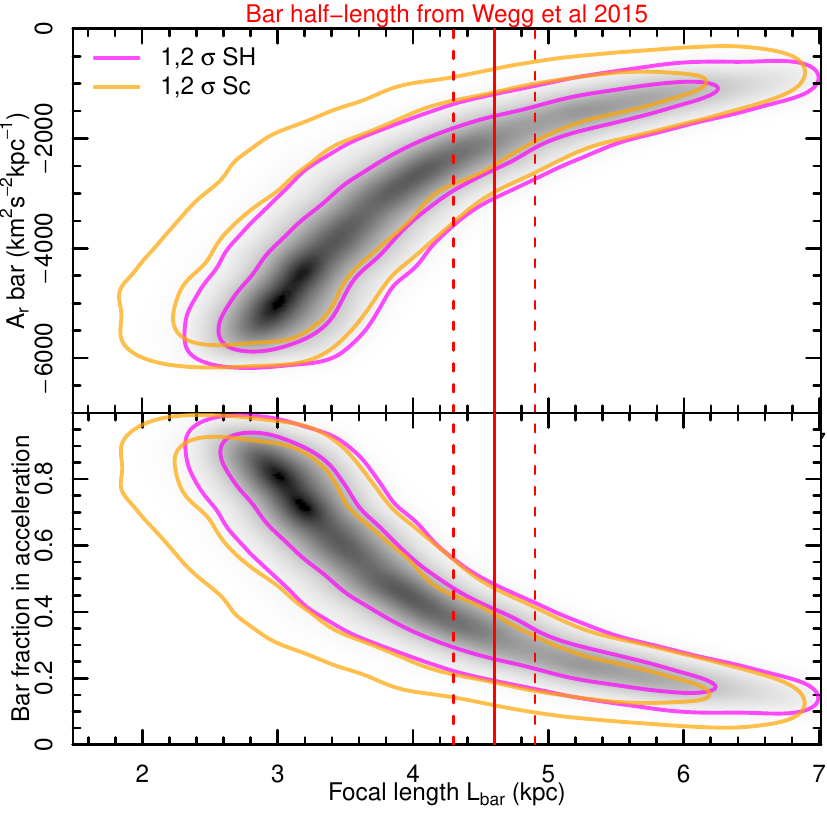}
    \caption{The \textit{top panel} depicts the degeneracy between the length of the bar and the acceleration from it. The \textit{bottom panel} shows the fraction of bar acceleration. The degeneracy can be broken when additional information such as bar length is used. We used the bar length value from \citet{Wegg:2015}. The contours show $1\sigma$ and $1\sigma$ confidence intervals for the  \citet{Khalatyan:2019} (SH) and \citet{Schonrich:2019} (Sc) datasets. }
    \label{fig:bar_focal}
\end{figure}

When using accelerations in the form of Eqs.~\eqref{eq:focus_eq1}--\eqref{eq:focus_eq3}, substantial correlations exist in the modelled posterior samples (e.g. between $A_r$ and $A_{r,{\rm bar}}$). The largest correlation coefficient was found to be $1.0$ between $A_r$ and $A_{r,{\rm bar}}$ (see Fig.~\ref{fig:ar_vs_ar_bar}). We also found a strong correlation (correlation coefficient  0.85) between bar acceleration/total acceleration and its focal length as shown in Fig.~\ref{fig:bar_focal}. The rest of the correlations were much weaker, and the only other noticeable correlation was between $A_{R}$ and $A_{R,{\rm bar}}$ and the radial acceleration derivative $A_{R,R}$ (correlation coefficient $0.26$). 

Currently we have only used the SN region to disentangle the bar parameters; hence, we were able to determine only some of the degenerate values. We were also able to determine that the sum of the radial acceleration due to the bar and axisymmetric components is constant, as seen in Fig.~\ref{fig:ar_vs_ar_bar}. Hence, if we know one then we can easily estimate the other. 

Based on the results from Sect.~\ref{sec:mustbebar}, we modelled the tangential acceleration value $A_y$. From this, we can calculate the $z$-component of the torque caused by the bar per solar mass using: 
$$T_z = A_yR_{\rm GC} \approx \{ 2450^{+420}_{-480} ({\rm SH}),\,2390^{+440}_{-660}({\rm Sc})\}\,{\rm km^2\,s^{-2}} $$
or 
$$ T_z \approx 2470\,{\rm km\,s^{-1}\,kpc\,Gyr^{-1}} .$$ 
Here $R_{\rm GC}$ is our physical distance from the Galactic centre (instead of the modelled radius of curvature of equipotential curves $R_\odot$). 
The degeneracy is because a long bar closer to us and a short massive bar engender the same force, making it difficult to distinguish the two scenarios.
This is seen in Fig.~\ref{fig:bar_focal} where the acceleration value due to the bar and the fraction of acceleration caused by the bar as a function of the length of the bar are given. The degeneracy can be broken when we fix the length of the bar from independent measurements. As an example, \citet{Wegg:2015} estimated that the half-length of the bar is $4.6\pm0.3\,{\rm kpc}$. If we fix this value as $L_{\rm bar}$, it gives the fraction of acceleration caused by the bar as about a third: $0.34\pm0.07$ in the case of the SH catalogue and  $0.29\pm0.09$ in the case of the Sc catalogue.

\section{Discussion and conclusion}

\subsection{Validity of the sample construction}\label{sec:test_fluxlim}

To test how well the approach described in Sect.~\ref{sec:impl_fluxlim} is able to cope with flux-limited data, we applied our model to two sets: a flux-limited sample and a volume-limited sample. The volume-limited approach was tested with simulation data in \citet{Kipper:2019} and was found to be consistent. Since the results for the flux-limited sample agreed well with those of the volume-limited sample, we are confident that our model is very suitable for the current case. 

The acceleration components used for this test are in the axisymmetric form of Eqs.~\eqref{eq:form_mustbebar3} -- \eqref{eq:form_mustbebar4}. The geometry of the region was biaxial ellipsoid in the form of Eq.~\eqref{eq:ell}, but its selection and modelling had some differences: the sample was limited by the absolute $J$-magnitude value of $1.2^{\rm m}$, {yielding} $54\,819$ stars. The grid was constructed by using $70$ random sample points, and fitting was done {eight} times to include the uncertainty from the gridding randomness. 

The results of the test calculations for volume- and flux-limited selections from the Sc catalogue are given in Table \ref{tab:mustbebar}, where calculated acceleration components are given with labels `Sc, vol' and `Sc, flux'. The most interesting acceleration components are radial acceleration $a_R$ and vertical acceleration $a_z$ (see their main parameters $A_R$ and $A_z$). For the volume-limited sample $A_R = -6128 \pm 199\,{\rm km^2s^{-2}kpc^{-1}}$ and $A_z = 183 \pm 106\,{\rm km^2s^{-2}kpc^{-1}}$, for the flux-limited sample $A_R = -6181 \pm 82\,{\rm km^2s^{-2}kpc^{-1}}$ and $A_z = 203 \pm 48\,{\rm km^2s^{-2}kpc^{-1}}$. The smaller errors in the flux-limited sample are due to the larger data sample. The results are clearly consistent and we may conclude that our approach to cope from here on with the flux-limited sample, described in Sect.~\ref{sec:impl_fluxlim}, is valid.

\subsection{Time dependence of acceleration due to the bar}\label{sec:time}

During calculations of stellar orbits (see selected analytical forms for acceleration) we assume that accelerations do not have an explicit time dependence. However, it is known that about a quarter of galaxies contain a more or less prominent bar \citep{Cheung:2013}; in the case of the MW a central bar was introduced by \citet{Devaucouleurs:1964}. A rotating bar would violate this assumption of our modelling. 

To test how much a bar would influence our results, we used the same simulation (the barred one) from \citet{Garbari:2011} as was used in \citet{Kipper:2019}. We selected a region close to the solar radius, with an angle between the major axis of the bar and direction to the centre of the region $\approx30^\circ$ and fitted the acceleration components \eqref{eq:form_cart_1} -- \eqref{eq:form_cart_3} (Taylor expansion) with our model.
% at that region. 
We expect that including the tangential component of the acceleration due to the bar which is changing in time would give us a somewhat wrong acceleration direction. We found that the acceleration vector was directed away by $3.27\pm2.23^\circ$ from the Galactic centre. The corresponding true angle calculated directly from the simulation gravitational potential was $2.21^\circ$. The difference between the true and calculated values is smaller than the $1\sigma$ error of the calculated value. Hence the effect of the time dependence of the bar in this case is {
not so significant}.

Another approach to estimate the effect of a time dependent gravitational potential is to use available data from the literature. The average angular speed of the bar is about $\sim 40\,{\rm km\,s^{-1}kpc^{-1}}$, although there is significant uncertainty in this value (see Sect.~\ref{sec:introduction}). The angular speed of the Sun is $\simeq 30 \,{\rm km\,s^{-1}kpc^{-1}}$\citep{BlandHawthorn:2016}. The strong similarity between these two values suggests that the potential of the bar near the Sun does not change very fast. A hypothetical star moving with a speed of $220\,{\rm km\,s^{-1}}$ passes the half box-size distance of $0.5$~kpc within $\sim 2.3$~Myr. Within this time, the angle of the bar orientation with respect to our comoving location changes only by $2.3\,\mathrm{Myr}\,\times\,(40-30)\,\mathrm{km\,s^{-1}kpc^{-1}} \approx 1.4^\circ$. If we assume that the angle between us and the major axis of the bar is $\theta_0 = 30^\circ$ and the 'tip of the bar' is about $L=5$~kpc away from the centre of the Galaxy\footnote{We make an approximation that the mass of the bar is a point mass at the tip of the bar. This gives an upper limit for the bar influence.} then our distance to the tip of the bar is  
\begin{equation}
    w^2 = L^2 + R_\odot^2 - 2LR_\odot\cos{\theta_0}.
\end{equation}
The force due to the bar changes within $2.3$~Myr maximally by
\begin{equation}
    \frac{\Delta F}{F} = \frac{1}{F}\, \frac{{\rm d} F}{{\rm d} w} \, \frac{{\rm d} w }{{\rm d} \theta_0 }\, \frac{{\rm d} \theta_0}{{\rm d} t}\Delta t \approx
    \frac{2LR_\odot\sin\theta_0}{w^2}\Delta \theta_0 \approx 0.046.
\end{equation}
Hence, the force due to the bar changes by about $5$ per cent if the force due to the bar is not steeper than $\propto w^{-2}$ and the bar dominates the potential. If it does not, then the result must be multiplied by the acceleration fraction of the bar. This can be considered as a component of systematic uncertainty. 
While calculating the orbital arcs for stars within the selected region, the time dependence of accelerations due to the bar has quite a small effect. Therefore in the current study we ignore this effect. 

\subsection{Influence of uncertainties in input data}

The input to this modelling does not include uncertainties. 
The resulting uncertainties are statistical in nature and include only sampling errors seen from the likelihood equation (7) of \citet{Kipper:2019}.
In order to see how {observational} uncertainties influence our results, we randomized phase space coordinates of stars according to their uncertainties, reconstructed the selection sample as described in Sect.~\ref{sec:data_and_selection}, and remodelled the selected region. To  account for the randomness in this process, we modelled the SN $47$ times and combined the corresponding posterior distributions. The results of the calculations are shown in Table~\ref{tab:mustbebar} with the label `Sc, rnd' after the variable name. Comparing calculated accelerations with labels `Sc, rnd' and `Sc, flux', it is seen that randomization had very little effect on the results. We conclude that uncertainties can be ignored for this selection. 

Another source of error can be due to the gridding approach employed in this study (see Sect.~\ref{sec:impl_grid}). Since there is randomization, we must include the noise caused by it. We rerun each modelling eight times to include the source of noise. All of these runs had similar posterior distributions; hence we are certain that gridding does not introduce large artificial uncertainties. To include the gridding uncertainties, we combined the posterior distributions of randomized grid runs.

\subsection{Conclusions}
In this paper, we have applied the \textit{orbital arc method} \citep{Kipper:2019} to \textit{Gaia} DR2 and modelled the acceleration along the plane of the Galactic disc.  We approximated the acceleration in the solar neighbourhood with various functional forms and came to the following conclusions:
\begin{enumerate}
    \item There are very few systematic biases between the \textit{Gaia} DR2 datasets %based on  and
    compiled by \citet{Schonrich:2019} and \citet{Khalatyan:2019}. Both the datasets give consistent results.
    \item The distribution of axisymmetric gravitational acceleration does not account for the observed acceleration for the standard distance of the Sun from the Galactic centre $R_\odot\approx8.3\,{\rm kpc}$. The curvature of the isopotential lines is smaller than the standard $R_\odot$, implying that there is a component of the Galactic bar causing this acceleration.
    \item The acceleration vector in the solar neighbourhood is not directed towards the centre of the Galaxy. There is a significant component of the acceleration directed away from the Galactic centre. We propose that this is caused by the massive central bar. Based on our model, we calculate the torque to be  $\sim 2400\,{\rm km^2\,s^{-2}}$ per solar mass. 
    \item Based on the assumption that isopotential surfaces of the bar are confocal ellipses, we estimate that about one third of the total acceleration in the {solar neighbourhood} is caused by the bar. In this computation we use the estimate of the length of the bar of \citet{Wegg:2015}. 
    \item Finally, using our model, we estimated the mass of the bar as $(1.6\pm0.3)\times10^{10}\,\mathrm{M}_\odot$, using the density distribution parameters from \citet{Wegg:2015}. 
\end{enumerate}

\section*{Acknowledgements}
We thank the referee for helpful comments and suggestions. 
We thank the StarHorse core team (F. Anders, A. Queiroz , B. Santiago, A. Kalathyan, C. Chiappini) for providing their data, and G. Monari for helpful comments about the paper. 
This work was supported by institutional research funding \mbox{IUT26-2}, \mbox{IUT40-2} and \mbox{PUTJD907} of the Estonian Ministry of Education and Research. We acknowledge the support by the Centre of Excellence ``Dark side of the Universe'' (TK133) and by the grant MOBTP86 financed by the European Union through the European Regional Development Fund. 
This work has made use of data from the European Space Agency (ESA) mission {\it Gaia} (\url{https://www.cosmos.esa.int/gaia}), processed by the {\it Gaia} Data Processing and Analysis Consortium (DPAC,
\url{https://www.cosmos.esa.int/web/gaia/dpac/consortium}). Funding for the DPAC has been provided by national institutions, in particular the institutions participating in the {\it Gaia} Multilateral Agreement. This publication makes use of data products from the Two Micron All Sky Survey, which is a joint project of the University of Massachusetts and the Infrared Processing and Analysis Center/California Institute of Technology, funded by the National Aeronautics and Space Administration and the National Science Foundation.

\bibliographystyle{mnras}
\bibliography{gaia_dyn} 

%%%%%%%%%%%%%%%%%%%%%%%%%%%%%%%%%%%%%%%%%%%%%%%%%%

\appendix
\section{Tables}
\begin{table*}
    \centering
    \caption{
    The modelling of the acceleration function described with Eqs.~\eqref{eq:form_cart_1}-\eqref{eq:form_cart_3} and using datasets from \citet{Schonrich:2019} (Sc) or \citet{Khalatyan:2019} (SH). We use acceleration units of ${\rm km^2\,s^{-2}\,kpc^{-1}}$, which differ from the more intuitive ${\rm km\,s^{-1}\,Gyr^{-1}}$ by about 2 per cent. The values of $P$ represent quantiles of the posterior distribution.
    }
    \label{tab:cartesian}
    \begin{tabular}{ll|ccccccc}
    \hline
        Variable & Unit & $P  = 0.02$ & $P = 0.16$ & Median & $P = 0.84$ & $P = 0.98$ & Lower prior limit & Higher prior limit\\
        \hline
$A_x$ (SH) &  km$^2\,$s$^{-2}$kpc$^{-1}$  &  6109.43  &  6178.27  &  6250.33  &  6382.74  &  6468.32  &  -10000  &  10000 \\ 
$A_x$ (Sc) &  km$^2\,$s$^{-2}$kpc$^{-1}$  &  6026.12  &  6102.47  &  6195.79  &  6327.5  &  6420.91  &  -10000  &  10000 \\ 
$A_y$ (SH) &  km$^2\,$s$^{-2}$kpc$^{-1}$  &  222.24  &  259.31  &  306.37  &  385.33  &  445.89  &  -10000  &  10000 \\ 
$A_y$ (Sc) &  km$^2\,$s$^{-2}$kpc$^{-1}$  &  189.42  &  238.87  &  283.55  &  339.59  &  412.18  &  -10000  &  10000 \\ 
$A_z$ (SH) &  km$^2\,$s$^{-2}$kpc$^{-1}$  &  138.96  &  175.95  &  211.06  &  254.66  &  295.01  &  -5000  &  5000 \\ 
$A_z$ (Sc) &  km$^2\,$s$^{-2}$kpc$^{-1}$  &  95.85  &  135.28  &  186.81  &  245.17  &  286.13  &  -5000  &  5000 \\ 
$A_{x,x}$ (SH) &   km$^2\,$s$^{-2}$kpc$^{-2}$  &  252.69  &  658.7  &  1152.87  &  1764.86  &  2306.3  &  -3000  &  5000 \\ 
$A_{x,x}$ (Sc) &   km$^2\,$s$^{-2}$kpc$^{-2}$  &  -498  &  318.77  &  1109.62  &  1631.22  &  2088.79  &  -3000  &  5000 \\ 
$A_{x,y}$ (SH) &   km$^2\,$s$^{-2}$kpc$^{-2}$  &  -34.2  &  853.24  &  1867.96  &  2920.27  &  3606.84  &  -4000  &  4000 \\ 
$A_{x,y}$ (Sc) &   km$^2\,$s$^{-2}$kpc$^{-2}$  &  -479.88  &  494.3  &  1511.29  &  2379.45  &  3128.42  &  -4000  &  4000 \\ 
$A_{x,z}$ (SH) &   km$^2\,$s$^{-2}$kpc$^{-2}$  &  -1948.43  &  -1760.52  &  -1406.23  &  -860.97  &  -171.53  &  -2000  &  2000 \\ 
$A_{x,z}$ (Sc) &   km$^2\,$s$^{-2}$kpc$^{-2}$  &  -1945.27  &  -1748.13  &  -1339.24  &  -725.73  &  2.09  &  -2000  &  2000 \\ 
$A_{y,x}$ (SH) &   km$^2\,$s$^{-2}$kpc$^{-2}$  &  -702.07  &  -424.69  &  -59.88  &  253.95  &  538.2  &  -2000  &  2000 \\ 
$A_{y,x}$ (Sc) &   km$^2\,$s$^{-2}$kpc$^{-2}$  &  -541.84  &  -283.02  &  -28.31  &  275.48  &  640.52  &  -2000  &  2000 \\ 
$A_{y,y}$ (SH) &   km$^2\,$s$^{-2}$kpc$^{-2}$  &  -3505.68  &  -2862.34  &  -2148.94  &  -1487.17  &  -935.21  &  -5000  &  2000 \\ 
$A_{y,y}$ (Sc) &   km$^2\,$s$^{-2}$kpc$^{-2}$  &  -3662.39  &  -2914.14  &  -2154.12  &  -1566.75  &  -933.65  &  -5000  &  2000 \\ 
$A_{y,z}$ (SH) &   km$^2\,$s$^{-2}$kpc$^{-2}$  &  -1975.02  &  -1885.74  &  -1696.72  &  -1372.29  &  -860.07  &  -2000  &  2000 \\ 
$A_{y,z}$ (Sc) &   km$^2\,$s$^{-2}$kpc$^{-2}$  &  -1957.05  &  -1817.54  &  -1463.88  &  -864.12  &  -129.38  &  -2000  &  2000 \\ 
$A_{z,x}$ (SH) &   km$^2\,$s$^{-2}$kpc$^{-2}$  &  -494.93  &  -145.76  &  155.8  &  428.63  &  700.61  &  -2000  &  2000 \\ 
$A_{z,x}$ (Sc) &   km$^2\,$s$^{-2}$kpc$^{-2}$  &  -48.17  &  184.64  &  429.13  &  669.57  &  906.36  &  -2000  &  2000 \\ 
$A_{z,y}$ (SH) &   km$^2\,$s$^{-2}$kpc$^{-2}$  &  -636.93  &  -50.74  &  562.24  &  1220.58  &  1772.55  &  -2000  &  2000 \\ 
$A_{z,y}$ (Sc) &   km$^2\,$s$^{-2}$kpc$^{-2}$  &  -304.58  &  102.45  &  501.94  &  893  &  1284.27  &  -2000  &  2000 \\ 
$A_{z,z}$ (SH) &   km$^2\,$s$^{-2}$kpc$^{-2}$  &  -3066.65  &  -2536.04  &  -1857.43  &  -1224.14  &  -664.61  &  -6000  &  0 \\ 
$A_{z,z}$ (Sc) &   km$^2\,$s$^{-2}$kpc$^{-2}$  &  -3556.61  &  -2789.1  &  -1651.53  &  -1081.78  &  -588.78  &  -6000  &  0 \\ 
\hline
    \end{tabular}
    
\end{table*}
\begin{table*}
    \centering
    \caption{
    The modelling of the acceleration function aiming to describe an  axisymmetric disc with a possible tangential component using equations \eqref{eq:form_mustbebar1} - \eqref{eq:form_mustbebar5}     and using datasets from \citet{Schonrich:2019} (Sc) or \citet{Khalatyan:2019} (SH). The extra denotations after the variable name show specifics of the modelling: 'flux' denotes that the sample was flux-limited and 'vol' volume-limited; 'rnd' had phase space values randomized according to observational uncertainties. {In the case of the random sample, the posterior distribution is averaged over $47$ different runs. } The volume-limited sample fit was done with about a tenth of the number of data, which is the cause of reduced accuracy and precision.    
    }
    \label{tab:mustbebar}
    \begin{tabular}{ll|ccccccc}
    Variable & Unit & $P  = 0.02$ & $P = 0.16$ & Median & $P = 0.84$ & $P = 0.98$ & Lower prior limit & Higher prior limit\\
    \hline
$A_{R}$ (Sc, vol) &  km$^2$\,s$^{-2}$kpc$^{-1}$ & -6495.2 & -6328.1 & -6127.9 & -5942.7 & -5785.8 & -15000 & 15000\\ 
$A_{R}$ (Sc, flux) &  km$^2$\,s$^{-2}$kpc$^{-1}$ & -6466.8 & -6300.4 & -6181.2 & -6110.2 & -6043.4 & -15000 & 15000\\ 
$A_{R}$ (SH, flux) &  km$^2$\,s$^{-2}$kpc$^{-1}$ & -6367.9 & -6294.8 & -6214 & -6135.1 & -6062.6 & -15000 & 15000\\ 
$A_{R}$ {(Sc, rnd)} &  km$^2$\,s$^{-2}$kpc$^{-1}$ & -6656.2 & -6517.4 & -6391.2 & -6284.1 & -6188.4 & -15000 & 15000\\ 
$R_\odot$ (Sc, vol) &  kpc & 1.6 & 2.1 & 3.6 & 9.9 & 17.2 & 0.1 & 20\\ 
$R_\odot$ (Sc, flux) &  kpc & 2 & 2.4 & 3.2 & 5 & 12.1 & 0.1 & 20\\ 
$R_\odot$ (SH, flux) &  kpc & 1.7 & 2.2 & 3.4 & 6.3 & 14.6 & 0.1 & 20\\ 
$R_\odot$ {(Sc, rnd)} &  kpc & 1.5 & 1.8 & 2.2 & 3.2 & 6.6 & 0.1 & 20\\ 
$A_{z}$ (Sc, vol) &  km$^2$\,s$^{-2}$kpc$^{-1}$ & 3.5 & 90.6 & 183.5 & 300.3 & 391 & -5000 & 5000\\ 
$A_{z}$ (Sc, flux) &  km$^2$\,s$^{-2}$kpc$^{-1}$ & 96.7 & 151.6 & 203.5 & 249.2 & 287.7 & -5000 & 5000\\ 
$A_{z}$ (SH, flux) &  km$^2$\,s$^{-2}$kpc$^{-1}$ & 114.7 & 152.8 & 195.5 & 239.3 & 280.9 & -5000 & 5000\\ 
$A_{z}$ {(Sc, rnd)} &  km$^2$\,s$^{-2}$kpc$^{-1}$ & 105.7 & 155.8 & 205 & 256.4 & 299.6 & -5000 & 5000\\ 
$A_{z,z}$ (Sc, vol) &  km$^2$\,s$^{-2}$kpc$^{-2}$ & -6815.1 & -5339.3 & -3731.1 & -1644.7 & -253.7 & -8000 & 0\\ 
$A_{z,z}$ (Sc, flux) &  km$^2$\,s$^{-2}$kpc$^{-2}$ & -3590.3 & -2872.3 & -2083.4 & -1427.1 & -755.2 & -8000 & 0\\ 
$A_{z,z}$ (SH, flux) &  km$^2$\,s$^{-2}$kpc$^{-2}$ & -3042.2 & -2285.3 & -1512 & -786.5 & -262.2 & -8000 & 0\\ 
$A_{z,z}$ {(Sc, rnd)} &  km$^2$\,s$^{-2}$kpc$^{-2}$ & -3478.4 & -2701.8 & -1866.7 & -1088.7 & -428.3 & -8000 & 0\\ 
$A_{z,R}$ (Sc, vol) &  km$^2$\,s$^{-2}$kpc$^{-2}$ & -1919.5 & -1542 & -991.9 & -232 & 282.9 & -5000 & 5000\\ 
$A_{z,R}$ (Sc, flux) &  km$^2$\,s$^{-2}$kpc$^{-2}$ & -742.1 & -446.4 & -145.4 & 176.9 & 473 & -5000 & 5000\\ 
$A_{z,R}$ (SH, flux) &  km$^2$\,s$^{-2}$kpc$^{-2}$ & -817.1 & -558.7 & -299.3 & -1.2 & 311.4 & -5000 & 5000\\ 
$A_{z,R}$ {(Sc, rnd)} &  km$^2$\,s$^{-2}$kpc$^{-2}$ & -949.9 & -653.9 & -373.8 & -43.5 & 320.3 & -5000 & 5000\\ 
$A_{z,Rz}$ (Sc, vol) &  km$^2$\,s$^{-2}$kpc$^{-3}$ & -4627.4 & -3272.8 & -104.6 & 3125.1 & 4569 & -5000 & 5000\\ 
$A_{z,Rz}$ (Sc, flux) &  km$^2$\,s$^{-2}$kpc$^{-3}$ & -4702.6 & -3640.8 & -1332.2 & 1973.6 & 4105.8 & -5000 & 5000\\ 
$A_{z,Rz}$ (SH, flux) &  km$^2$\,s$^{-2}$kpc$^{-3}$ & -4814.5 & -3985.5 & -1707.9 & 2408 & 4426 & -5000 & 5000\\ 
$A_{z,Rz}$ {(Sc, rnd)} &  km$^2$\,s$^{-2}$kpc$^{-3}$ & -4407.6 & -2608.8 & 788.9 & 3471.8 & 4683.1 & -5000 & 5000\\ 
$A_{R,R}$ (Sc, vol) &  km$^2$\,s$^{-2}$kpc$^{-2}$ & -2342.4 & -938.9 & 909.7 & 2490.4 & 3444 & -4000 & 4000\\ 
$A_{R,R}$ (Sc, flux) &  km$^2$\,s$^{-2}$kpc$^{-2}$ & -2591.1 & -1809.7 & -995.1 & -102 & 975.3 & -4000 & 4000\\ 
$A_{R,R}$ (SH, flux) &  km$^2$\,s$^{-2}$kpc$^{-2}$ & -2726.1 & -1666.2 & -500.3 & 531 & 1372.3 & -4000 & 4000\\ 
$A_{R,R}$ {(Sc, rnd)} &  km$^2$\,s$^{-2}$kpc$^{-2}$ & -2591.9 & -1721.2 & -763.8 & 455.6 & 1799.6 & -4000 & 4000\\ 
$A_{y}$ (Sc, vol) &  km$^2$\,s$^{-2}$kpc$^{-1}$ & -345.2 & -226.3 & -65 & 67.7 & 177.8 & -3000 & 3000\\ 
$A_{y}$ (Sc, flux) &  km$^2$\,s$^{-2}$kpc$^{-1}$ & 157.2 & 208.5 & 288.5 & 341.5 & 385.5 & -3000 & 3000\\ 
$A_{y}$ (SH, flux) &  km$^2$\,s$^{-2}$kpc$^{-1}$ & 159.3 & 237.3 & 295.4 & 345.9 & 400.4 & -3000 & 3000\\ 
$A_{y}$ {(Sc, rnd)} &  km$^2$\,s$^{-2}$kpc$^{-1}$ & 134.6 & 187.3 & 247.5 & 309.6 & 393 & -3000 & 3000\\ 
    \end{tabular}
\end{table*}
\begin{table*}
    \centering
    \caption{The modelling of the acceleration function  \eqref{eq:focus_eq1}--\eqref{eq:focus_eq3} aiming to describe the sum of the axisymmetric and confocal bar components. The datasets used from  \citet{Schonrich:2019} and \citet{Khalatyan:2019} are abbreviated as Sc and SH.}
    \label{tab:my_label}
    \begin{tabular}{ll|ccccccc}
    \hline
    Variable & Unit & $P  = 0.02$ & $P = 0.16$ & Median & $P = 0.84$ & $P = 0.98$ & Lower prior limit & Higher prior limit\\
    \hline
$A_{R}$ (SH)  &  km$^2$\,s$^{-2}$kpc$^{-1}$  &  -5347.36  &  -4713.55  &  -3107.82  &  -1291.7  &  -513.77  &  -10000  &  0 \\ 
$A_{R}$ (Sc)  &  km$^2$\,s$^{-2}$kpc$^{-1}$  &  -5508.89  &  -4889.42  &  -3195.73  &  -1344.89  &  -513.04  &  -10000  &  0 \\ 
$A_{R,R}$ (SH)  &  km$^2$\,s$^{-2}$kpc$^{-2}$  &  234.31  &  766.59  &  1248.09  &  1717.45  &  2202.51  &  -5000  &  5000 \\ 
$A_{R,R}$ (Sc)  &  km$^2$\,s$^{-2}$kpc$^{-2}$  &  -739.71  &  164.14  &  1195.15  &  1966.9  &  2599.58  &  -5000  &  5000 \\ 
$A_z$ (SH)  &  km$^2$\,s$^{-2}$kpc$^{-1}$  &  111.05  &  150.27  &  202.37  &  248.64  &  287.65  &  -5000  &  5000 \\ 
$A_z$ (Sc)  &  km$^2$\,s$^{-2}$kpc$^{-1}$  &  141.71  &  178.88  &  216.77  &  256.01  &  293.62  &  -5000  &  5000 \\ 
$A_{z,z}$ (SH)  &  km$^2$\,s$^{-2}$kpc$^{-2}$  &  -3331.1  &  -2737.24  &  -2135.29  &  -1479.55  &  -897.58  &  -8000  &  0 \\ 
$A_{z,z}$ (Sc)  &  km$^2$\,s$^{-2}$kpc$^{-2}$  &  -3090.27  &  -2461.84  &  -1856.71  &  -1156  &  -582.57  &  -8000  &  0 \\ 
$A_{z,R}$ (SH)  &  km$^2$\,s$^{-2}$kpc$^{-2}$   &  -873.7  &  -597.21  &  -286.73  &  -6.01  &  237.43  &  -5000  &  5000 \\ 
$A_{z,R}$ (Sc)  &  km$^2$\,s$^{-2}$kpc$^{-2}$   &  -933.1  &  -636.23  &  -345.07  &  -83.41  &  154.43  &  -5000  &  5000 \\ 
$A_{z,Rz}$ (SH)  &  km$^2$\,s$^{-2}$kpc$^{-3}$  &  -4352.62  &  -2373.53  &  1429.96  &  3747.56  &  4718.91  &  -5000  &  5000 \\ 
$A_{z,Rz}$ (Sc)  &  km$^2$\,s$^{-2}$kpc$^{-3}$  &  -4341.86  &  -2571.03  &  399.29  &  3137.3  &  4610.42  &  -5000  &  5000 \\ 
$A_{R, {\rm bar}}$ (SH)  &  km$^2$\,s$^{-2}$kpc$^{-1}$  &  -5761.45  &  -4984.16  &  -3168.97  &  -1562.17  &  -952.39  &  6000  &  -6000 \\ 
$A_{R, {\rm bar}}$ (Sc)  &  km$^2$\,s$^{-2}$kpc$^{-1}$  &  -5738.29  &  -4899.51  &  -3050.55  &  -1367.63  &  -737.13  &  6000  &  -6000 \\ 
$L_{\rm bar}$ (SH)  &  kpc  &  2.62  &  3.03  &  3.76  &  5.18  &  6.4  &  0.1  &  7 \\ 
$L_{\rm bar}$ (Sc)  &  kpc  &  2.16  &  2.72  &  3.54  &  5.02  &  6.32  &  0.1  &  7 \\ 
\hline
    \end{tabular}

\end{table*}

\bsp
\label{lastpage}
\end{document}